\documentclass[lettersize,journal]{IEEEtran}
\usepackage{amsmath,amsfonts}
\usepackage{algorithmic}
\usepackage{algorithm}
\usepackage{array}
\usepackage[caption=false,font=normalsize,labelfont=sf,textfont=sf]{subfig}
\usepackage{textcomp}
\usepackage{stfloats}
\usepackage{url}
\usepackage{verbatim}
\usepackage{graphicx}
\usepackage{cite}
\usepackage{caption} 
\usepackage{color} 
\usepackage{adjustbox} 
\usepackage{xcolor}  
\begin{document}

\title{Low-Latency Federated Fine-Tuning for Large Language Models Over Wireless Networks}

\author{Zhiwen Pang,
        Kang Wei,
        Long Shi,
        Zhe Wang,
        Jun Li,~\IEEEmembership{Fellow,~IEEE,}
        and Feng Shu
\thanks{Zhiwen Pang and Long Shi are with the School of Electronic and Optical Engineering, Nanjing University of Science and Technology, Nanjing 210094, China (e-mails:$\lbrace$zhiwenpang, longshi$\rbrace$@njust.edu.cn). 
Kang Wei is with the School of Cyber Science and Engineering, Southeast University, Nanjing 211189, China (e-mail: kang.wei@seu.edu.cn). 
Zhe Wang is with the School of Computer Science and Engineering, Nanjing University of Science and Technology, Nanjing 210094, China (e-mail: zwang@njust.edu.cn).
Jun Li is with the School of Information Science and Engineering, Southeast University, Nanjing 211189, China (e-mail: jun.li@seu.edu.cn).
Feng Shu is with the School of Information and Communication Engineering, Hainan University, Haikou 570228, China (e-mail: shufeng0101@163.com).
}}

\markboth{Journal of \LaTeX\ Class Files,~Vol.~14, No.~8, August~2021}%
{Shell \MakeLowercase{\textit{et al.}}: A Sample Article Using IEEEtran.cls for IEEE Journals}


\maketitle

\maketitle

\begin{abstract}
Recently, federated large language models (LLMs) have drawn significant attention thanks to coupled capabilities of LLMs and federated learning (FL) that address privacy concerns in collaborative fine-tuning. However, due to large-scale parameters of LLMs, existing federated LLM fine-tuning frameworks incur significant challenges in resource-constrained clients characterized by heterogeneous computing capabilities and random wireless channels. To address this issue, we propose a joint client-specific pruning and bandwidth allocation (JCPBA) framework for federated LLMs to improve the fine-tuning efficiency over the wireless networks. 
Specifically, we formulate a fine-tuning latency minimization problem by jointly optimizing pruning rates and bandwidth allocations.
Furthermore, we solve this optimization problem using a block coordinate descent method. 
Extensive experiments on the datasets of Yahoo Answers and GSM8K demonstrate that the proposed framework significantly reduces wall-clock fine-tuning time compared with state-of-the-art baselines and gains equal or lower test loss at the cost of lower computation and communication overhead.
\end{abstract}


\section{Introduction}
\IEEEPARstart{L}{arge} language models (LLMs) have recently achieved remarkable progress in diverse tasks such as question answering, logical reasoning, and generative content creation~\cite{Liu2023Towards}. 
The success of LLMs significantly relies on large datasets and high-performance computing, which poses challenges for many organizations, particularly around data privacy~\cite{zhu2026moe,shi2025stable}.
However, a large amount of high-quality private data is widely distributed among various data holders, which cannot be publicly shared due to privacy concerns, regulation restrictions, or other reasons~\cite{ref4}.
For example, such specific domain data, e.g., medical records and financial information, are often sensitive and cannot be easily centralized because of privacy concerns and regulatory constraints~\cite{Hu2025Federated, jiang2025towards}.
Federated learning (FL) provides a solution through a distributed training/fine-tuning paradigm in which participants collaborate without sharing raw data, only relying on exchanging only model updates, thereby mitigating privacy risks~\cite{ref6}.

Although integrating FL and LLMs offers a promising solution to collaboratively fine-tune a shared LLM~\cite{Wei2023Personalized,Zhu2025Trustworthy,deng2025federated}, two major challenges remain. First, the massive scale of LLMs imposes heavy computation and communication demands, which hinders deployment in realistic edge device environments~\cite{ref5}. Second, clients with limited and heterogeneous computing, communication, and memory resources often suffer from stragglers and large variance in per-round latency, which inevitably undermine the overall training efficiency~\cite{ref8}.

To address these challenges, prior studies have explored two main directions of fine-tune LLMs in the federated paradigm. 
The first approach, focusing on efficiency, integrates parameter-efficient fine-tuning (PEFT), e.g., low-rank adaptation (LoRA), and model compression methods into LLM fine-tuning to substantially reduce computation and communication overhead~\cite{ref1, ref9}. 
For instance, FedBiOT develops an emulator–adapter design that decomposes the LLM into two functionally distinct components. The emulator is constructed from early-to-middle layers to preserve the model’s general representation capability. In contrast, the adapter is formed by later layers to capture task- and domain-specific patterns. During federated fine-tuning, clients keep the emulator fixed and only optimize the adapter using local data, eliminating the need to update the full model. As a result, this design substantially reduces computational and communication overhead in federated settings while maintaining model performance~\cite{ref1}.
To further improve efficiency, FedQLoRA introduces a quantization-aware adapter that decouples quantization error from LoRA updates and alternates between optimizing the error-compensation term and the LoRA parameters, which helps reduce quantization bias and heterogeneity bias while remaining efficient~\cite{ref9}. 
Another direction is dedicated to resource management by jointly optimizing communication and computation efficiency~\cite{ref7}.
Efforts in this area, such as DEFT, leverage joint optimization of depth-aware block partitioning and bandwidth-aware allocation to minimize latency during local training; however, the block assignment mechanism is complex and requires effective inter-device coordination~\cite{ref2}.
Other methods integrate model splitting and device scheduling to reduce latency by skipping underperforming clients, thereby shrinking the parameter space~\cite{ref3, ref10}.

While existing methods have successfully reduced overall resource costs, they still struggle to minimize fine-tuning latency, especially when clients are constrained by limited computation resources and unreliable wireless transmissions.
To overcome this challenge, we propose a low-latency fine-tuning framework for federated LLM that optimizes dynamic pruning and bandwidth allocations to minimize fine-tuning latency across heterogeneous clients. 
The main contributions of this letter are as follows:
\begin{itemize}

\item We develop a joint client-specific pruning and bandwidth allocation (JCPBA) framework based on the emulator--adapter design. To address the client heterogeneity, the server assigns a client-specific pruning rate and bandwidth according to each client’s resource heterogeneity and channel randomness, partitions the pre-trained model into an emulator and an adapter, and applies structured pruning to the emulator before dispatch, while clients only update the adapter using local data.

\item  We formulate a fine-tuning latency minimization problem over pruning rates and bandwidth allocations under resource and convergence constraints. To solve this problem, we develop the joint client-specific pruning and bandwidth allocation algorithm by using a block coordinate descent (BCD) method.

\item  On the datasets of Yahoo Answers and GSM8K, JCPBA reduces wall-clock fine-tuning time by at least 40\% compared with state-of-the-art baselines, while achieving equal or lower test loss and reducing communication overhead by at least 46\%.

\end{itemize}

The remainder of this letter is organized as follows. Section II presents the system model. Section III formulates the fine-tuning latency minimization problem and its solution. Section IV presents experimental results. Finally, Section V concludes this paper.
\section{System Model}
\subsection{Federated LLM Fine-Tuning with Pruning}
\begin{figure}[!t] 
\centering
\includegraphics[width=0.45\textwidth]{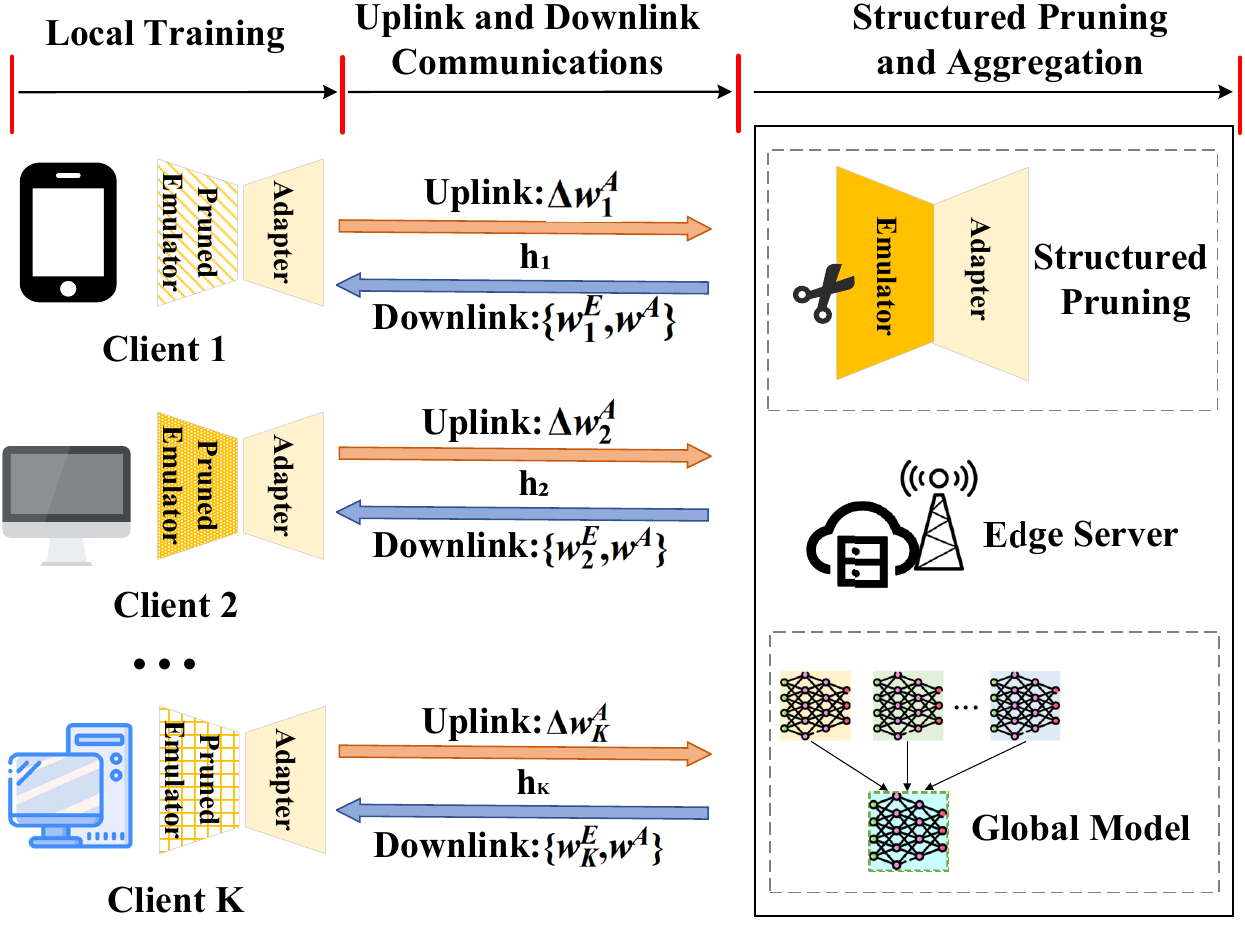}
\caption{The framework of federated LLM fine-tuning.}
\label{fig:1}
\end{figure}

As shown in Fig.~\ref{fig:1}, we consider a federated LLM fine-tuning framework over wireless networks that comprises a base station (BS) with an edge server and heterogeneous clients~\cite{zhu2025randomized}. The clients exhibit heterogeneity in computation capabilities and wireless channel conditions. Clients retain private local datasets and collaboratively fine-tune a pre-trained LLM through multiple rounds, aiming to minimize per-round latency without sacrificing the model performance.

This framework builds upon a centralized star topology, where all the clients communicate with the server over wireless channels. The server manages the global model by sequentially conducting model partitioning, model compression, pruning policy scheduling, bandwidth allocation, and global aggregation. Each client $k \in \mathcal{K}$ holds a private raw dataset $\mathcal{D}_k$ that is not transmitted and only returns adapter updates to the server for privacy preservation, where $\mathcal{K}$ is the set of client indices. Specifically, a federated fine-tuning round operates as follows:

\subsubsection{Model Dispatch} 
The server partitions the pre-trained LLM into an emulator with parameters $\boldsymbol{w}^E$ and an adapter with parameters $\boldsymbol{w}^A$~\cite{ref1}. 
To address the client heterogeneity, the server applies structured pruning to the emulator before transmitting the model to clients, by removing entire attention heads and multi-layer perceptron (MLP) neurons ~\cite{ref12}. Given the heterogeneous client conditions, the server assigns each client $k$ a client-specific pruning rate $\beta_k$ and bandwidth $B_k$, and then prunes the emulator parameters as
\begin{equation}
  \boldsymbol{w}^{E}_{k} = \mathcal{P}(\boldsymbol{w}^{E}, \beta_k),
  \label{eq:pruning}
\end{equation}
where $\mathcal{P}(\cdot)$ denotes a structured pruning function that removes attention heads and MLP neurons based on the pruning rate $\beta_k$. 
After pruning, the server sends a compressed model package: 
\begin{equation}
  M^{c}_{k} = \{\boldsymbol{w}^{E}_{k}, \boldsymbol{w}^{A}\},
  \label{eq:compressed_package}
\end{equation}
to client $k$ over the downlink channel.

\subsubsection{Local Fine-tuning}
Upon receiving the adapter parameters $\boldsymbol{w}^{A}$, client $k$ freezes the pruned emulator $\boldsymbol{w}^E_{k}$, updates the adapter parameters using its private data $\mathcal{D}_k$, and generates the locally updated adapter $\boldsymbol{w}^{A}_{k}$.

\subsubsection{Parameter Upload}
After local fine-tuning, client $k$ only uploads the adapter update (i.e., the parameter delta) to the server over the uplink channel:
\begin{equation}
  \Delta \boldsymbol{w}^{A}_{k} = \boldsymbol{w}^{A}_{k} - \boldsymbol{w}^{A},
  \label{eq:adapter_update}
\end{equation}
where $\Delta \boldsymbol{w}^{A}_{k}$ represents the parameter delta, namely the difference between the locally updated adapter parameters and the original adapter parameters.

\subsubsection{Update Aggregation}
The server aggregates all the updates by weighted FedAvg~\cite{Kang2020Reliable}:
\begin{equation}
  \boldsymbol{w}^A \leftarrow \boldsymbol{w}^A + 
  \sum_{k \in \mathcal{K}} \frac{|\mathcal{D}_k|}{\sum_{k'\in \mathcal{K}} |\mathcal{D}_{k'}|} \, \Delta \boldsymbol{w}^{A}_{k},
  \label{eq:fedavg}
\end{equation}
where $|\mathcal{D}_k|$ denotes the size (i.e., the number of samples) of the local dataset at client $k$. 
After aggregation, the server fine-tunes the emulator once per federated fine-tuning round with public data to maintain the general representation capability while aligning with task-specific patterns. Due to the abundant computational resources at the server, the server-side fine-tuning latency is negligible compared with that of the client-side, thereby not being included in the total fine-tuning latency.
 
Note that no raw data or emulator parameters are transmitted, which thereby reduces the exposure of sensitive information compared with transmitting full model updates.

\subsection{Computation and Communication Models}
In this subsection, we model the computation and communication latency of each client in a single federated fine-tuning round, which serves as the basis for the subsequent latency minimization problem.

\subsubsection{Computation Model}
In each round, the server dispatches the compressed model $M^{c}_{k}$, i.e., the pruned emulator and adapter, to all the clients, including the initialized parameters $\{ \boldsymbol{w}^{E}_{k}, \boldsymbol{w}^{A} \}$.
Each client \(k \in \mathcal{K}\) fine-tunes the parameters of the adapter \(\boldsymbol{w}_A\) using its private dataset \(\mathcal{D}_k\), while the parameters of the emulator \(\boldsymbol{w}^E_{k}\) remain frozen.

Since the emulator is pruned, its computation overhead depends on the pruning rate \(\beta_k\). 
Let $f_k$ denote the normalized computation speed of client $k$, which is a unitless indicator of its relative processing speed. Each local fine-tuning round consists of $M$ iterations, with each iteration processing a mini-batch of $N$ samples.
The computation latency per iteration at client $k$ is given by
\begin{equation}
d_k = a + e_0 (1 - \beta_k),
\label{eq:local_delay}
\end{equation}
where \( a \) denotes the computation cost of the adapter, which is fixed since the adapter structure remains unchanged over different iterations and is not subject to pruning, and \( e_0 \) is the computation cost of the unpruned emulator.
From (5), \( d_k \) is inversely correlated with \( \beta_k \), since a higher pruning rate reduces the computation cost of the emulator~\cite{ref2}. Using (5), the total latency of local fine-tuning at client \( k \) over \( M \) iterations is

\begin{equation}
T_{k,\beta_{k}}^{\text{comp}} = \frac{M  (a + e_0(1 - \beta_k))}{f_k}.
\label{eq:comp_latency}
\end{equation}

\subsubsection{Communication Model}
Each communication round refers to the phase in which the server transmits the compressed model to the clients, and the clients upload their updated adapter parameters back to the server. Each communication round follows a federated fine-tuning round, which includes the local model adaptation and update phases. Let \( B_k \) denote the bandwidth allocated for client \( k \). The downlink and uplink transmission rates between the server and client \( k \) are respectively:

\begin{equation}
R_k^{\downarrow} = B_k \log_2 \left( 1 + \frac{p_s H_k^{\downarrow}}{N_0} \right),  R_k^{\uparrow} = B_k \log_2 \left( 1 + \frac{p_k H_k^{\uparrow}}{N_0} \right),
\label{eq:spectral-eff}
\end{equation}
where \( p_s \) and \( p_k \) represent the transmit power of the server and client  respectively, \( H_k^{\downarrow} \) and \( H_k^{\uparrow} \) represent the downlink and uplink channel gains between the server and client \( k \) respectively, and \( N_0 \) is the white Gaussian noise power.

Notably, client \( k \) needs to download the compressed emulator \( \boldsymbol{w}^E_k \) and the adapter \( \boldsymbol{w}^A \) from the server, and only uploads the adapter update  \( \Delta \boldsymbol{w}^A_k \). In this context, the total communication latency of client \( k \) is given by

\begin{equation}
T_{k,\beta_{k}}^{\mathrm{comm}} = \frac{|\boldsymbol{w}^A| + (1 - \beta_k) |\boldsymbol{w}^E|}{R_k^{\downarrow}} + \frac{|\Delta \boldsymbol{w}^A_k|}{R_k^{\uparrow}},
\label{eq:comm-lat}
\end{equation}
where $|\cdot|$ represents the size (in bits) of the corresponding model parameters.

\section{Problem Formulation and Optimization}

The goal of this section is to minimize the overall fine-tuning latency under heterogeneity in client computation capability, memory capacity, and wireless channel states. To simplify the analysis, we reformulate the overall latency minimization problem as a per-round latency minimization problem to reduce the complexity to a single-round optimization.


From (6) and (8), the total latency for client \(k\) consists of local computation and communication. Collectively, the total latency per round for client \(k\) is given by
\begin{equation}
T_k = T_{k,\beta_k}^{\text{comp}} + T_{k,\beta_k}^{\text{comm}}.
\end{equation}


\begin{algorithm}[t]
\caption{JCPBA}
\label{algo:bcd}
\begin{algorithmic}[1]
\REQUIRE Number of clients \(K\), total bandwidth \(B\), pruning bounds \([\beta_{\min}, \beta_{\max}]\), convergence parameters \((\xi, \phi, \psi)\), stopping threshold $\epsilon$, maximum number of iterations $T_{\max}$.
\ENSURE Optimized pruning rates \(\{\beta_k\}\) and bandwidth allocations \(\{B_k\}\).
\STATE Initialize \(\beta_k \in [\beta_{\min}, \beta_{\max}]\), \(B_k = B/K\), and set iteration index \(t \gets 0\)
\STATE Initial objective \(T^{(0)} = \max_{k \in \mathcal{K}} T_k(\beta_k, B_k)\)
\REPEAT
    \STATE \textbf{Pruning Update:} Fix \(\{B_k\}\) and update \(\beta_k\) by
    \[
    \min_{\beta_k} \; T_k(\beta_k \mid B_k), \quad \text{s.t.} \quad \textbf{C3, C4, C5}.
    \]
    \STATE \textbf{Bandwidth Update:} Fix \(\{\beta_k\}\) and update \(\{B_k\}\) by 
    \[
    \min_{B_k} \; T_k(B_k \mid \beta_k), \quad \text{s.t.} \quad \textbf{C1, C2}.
    \]
    \STATE \(t \gets t + 1\)
    \STATE Compute \(T^{(t)} = \max_{k \in \mathcal{K}} T_k(\beta_k, B_k)\)
\UNTIL \(\left| T^{(t)} - T^{(t-1)} \right| < \epsilon\) or \(t \geq T_{\max}\)
\end{algorithmic}
\end{algorithm}
The objective is to jointly optimize the pruning rate \(\beta_k\) and bandwidth allocation \(B_k\) for each client \(k\), such that the maximum latency \(T\) is minimized.
Overall, we formulate the optimization problem as 
\begin{equation}
\begin{aligned}
\textbf{P1}:\,\, &\min_{\{\beta_k, B_k\}} \; T = \max_{k \in \mathcal{K}} T_k \\
&\text{s.t.}\quad\textbf{C1}–\textbf{C5},
\end{aligned}
\end{equation}
where \textbf{C1}–\textbf{C5} characterize the resource and convergence constraints of the federated fine-tuning process.
\begin{itemize}
  \item \textbf{C1 (Total bandwidth constraint):} The total bandwidth allocated to all the clients cannot exceed the server's available bandwidth \(B\), i.e., $\sum_{k \in \mathcal{K}} B_k \leq B$.
  \item \textbf{C2 (Non-negativity of client bandwidth):} The bandwidth allocated for each client must be non-negative, i.e., \(B_k \geq 0\).

  \item \textbf{C3 (Local memory constraint):} Each client stores the compressed model within its memory limit \(\hat{b}_k\), i.e., \( b(\beta_k) \leq \hat{b}_k \). 
  \item \textbf{C4 (Pruning rate feasibility):} Since excessive pruning degrades performance,  \(\beta_k\) is required to be within a feasible range, i.e., $\beta_{\min} \leq \beta_k \leq \beta_{\max}$.
  \item \textbf{C5 (Convergence constraint):} To ensure convergence, the effects of pruning and bandwidth allocation yield
  \[
  \xi + \frac{\phi}{K N} + \frac{\psi}{K} \sum_{k \in \mathcal{K}} \beta_k \leq \gamma_{\min},
  \]
  where \(\xi\), \(\phi\), and \(\psi\) are positive constants, and \(\gamma_{\min}\) bounds the gradient norm.

The convergence constraint \textbf{C5} in \textbf{P1}  is used to promote the convergence stability by bounding the gradient norms of federated LLM fine-tuning [15].
\end{itemize}
To solve this optimization, we adopt a classical BCD method that decomposes \textbf{P1} into two alternating subproblems: optimizing the pruning rates \(\{\beta_k\}\) and the inverse bandwidth allocations \(\{1/B_k\}\) respectively. The steps of BCD for solving \textbf{P1} are detailed in \textbf{Algorithm \ref{algo:bcd}}. In each iteration, one block of variables is optimized while the other is fixed. Specifically, in the \textbf{Pruning Update}, we optimize the pruning rate \(\beta_k\) while fixing the bandwidth \(\{B_k\}\), as pruning mainly affects computational cost and is constrained by \(\textbf{C3}\), \(\textbf{C4}\), and \(\textbf{C5}\). In the \textbf{Bandwidth Update}, we optimize the bandwidth \(\{B_k\}\) while fixing the pruning rates \(\{\beta_k\}\), under communication constraints \(\textbf{C1}\) and \(\textbf{C2}\). 

It is notable that JCPBA addresses heterogeneous computation and memory capacities by optimizing client-specific pruning rates and bandwidth allocations. The pruning rate is adjusted based on each client’s computation and memory capacity, allowing JCPBA to efficiently adapt to clients with varying resource constraints. This flexibility ensures that JCPBA can maintain more stable performance across heterogeneous environments than the existing related works.

Because the pruning update and the bandwidth update have linear complexity with respect to the number of clients in each iteration, the overall computational complexity of \textbf{Algorithm~1} is \(O(T_{\max} K)\), where \(K\) is the number of clients and \(T_{\max}\) is the number of iterations, respectively. Since \(T_{\max}\) is small in practice, the computational overhead of \textbf{Algorithm~1} is negligible compared with the client-side LLM fine-tuning cost.
\section{Experimental Results}
\subsection{Experiment Setup}
To evaluate the effectiveness of our method, we conduct a series of experiments on various datasets and models.
\subsubsection{Tasks and Datasets}
We use GPT-2 Medium as the pre-trained model for federated fine-tuning, and focus on two representative downstream tasks: (i) text classification over the Yahoo Answers dataset, and (ii) mathematical reasoning over the GSM8K dataset. For both tasks, clients only receive the compressed emulator and adapter, rather than the full LLM parameters. The server fine-tunes the emulator using publicly available data. Client data is independently and identically distributed across all clients.
\subsubsection{Fine-tuning Model}
We use GPT-2 Medium with 24 decoder layers, wherein layers 0-21 are designated as the emulator and layers 22-23 form the adapter. Client-side fine-tuning is restricted to the adapter, with the emulator remaining frozen. Each client only uploads the updated adapter parameters in each round.
We set the learning rate and batch size to \(1 \times 10^{-4}\) and 4, respectively.
\subsubsection{Communication Parameters}
We consider that a BS hosts the edge server and connects to 8 clients. The wireless links between the server and the clients are modeled as Rayleigh fading with a path-loss of \( 60 \)~dB.
 The server transmits at $10$~W, each client at $0.2$~W, and the total available communication bandwidth is $100$~MHz.
\subsubsection{Client Heterogeneity}
To reflect the client heterogeneity, we set the baseline computation speed to \( f_0 = 1~\mathrm{TFLOPS} \). For the \( k \)-th client, we randomly draw a factor \( \xi_k \) from a uniform distribution \( \mathcal{U}[0.5, 2] \) to scale the client's computation speed relative to the baseline \( f_0 \). The computation speed \( f_k \) of the \( k \)-th client is then calculated as \( f_k = \xi_k  f_0 \). The available client memory \( \hat{b}_k \) is independently sampled from a uniform distribution \( \mathcal{U}[4, 8] \) GB. 
\subsubsection{Baselines} To assess the gains of latency and accuracy, we compare JCPBA with baselines as follows:
\begin{itemize}
  \item \textbf{FedBiOT }\cite{ref1}: An emulator-adapter dual-level optimization paradigm where clients only update LoRA based adapter and the server aligns the emulator via distillation.
  \item \textbf{DEFT} ~\cite{ref2}: A depth-aware block allocation plus bandwidth allocation for joint communication and computation scheduling.
  \item \textbf{Split LoRA}~\cite{ref4}: A split-LoRA FL framework with Lyapunov-drift-based online client scheduling and bandwidth allocation.
  \item \textbf{UBFP:} The uniform bandwidth allocation and fixed pruning strategy.
\end{itemize}
\begin{figure}[t]
\centering
\includegraphics[width=3.5in]{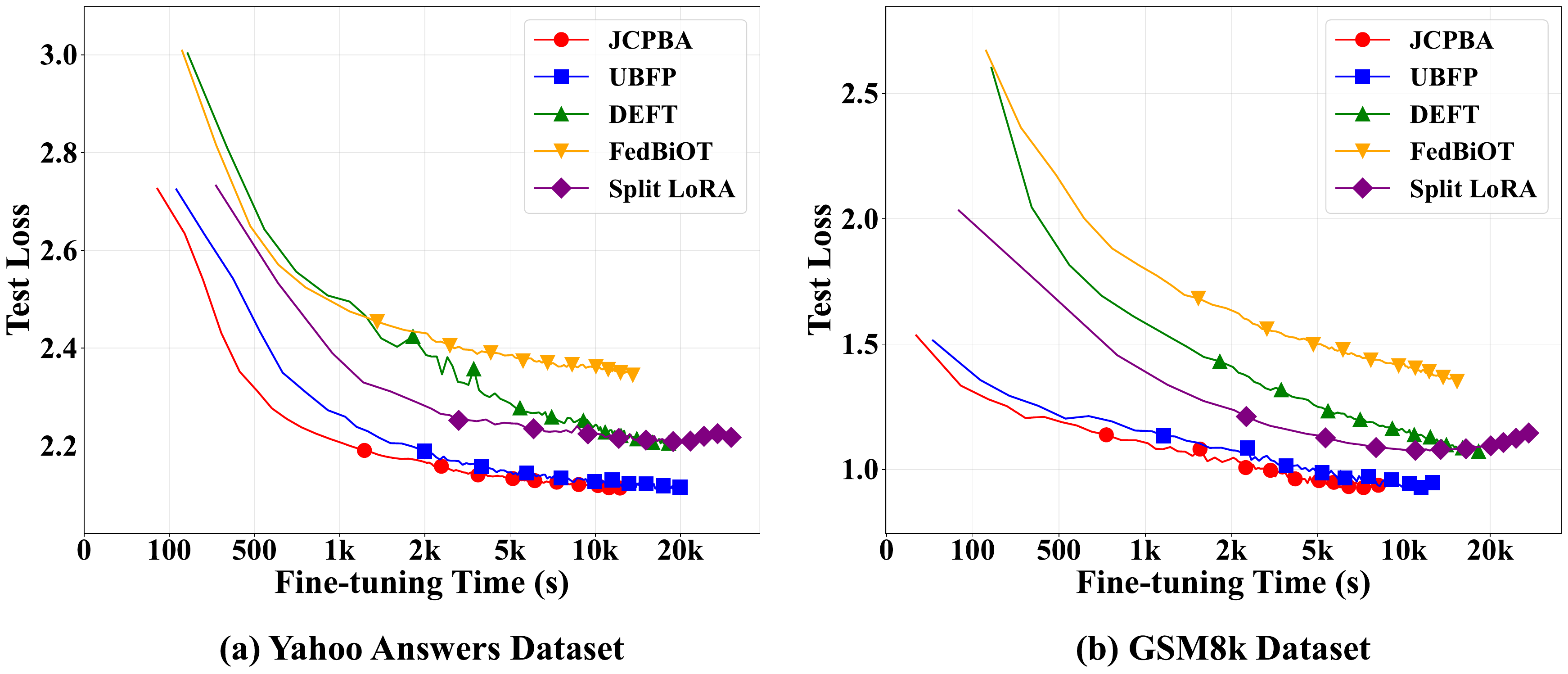}
\caption{The comparison of test loss on Yahoo Answers and GSM8K.}
\label{fig:results}
\end{figure}
\begin{figure}[t]
\centering
\includegraphics[width=3.5in]{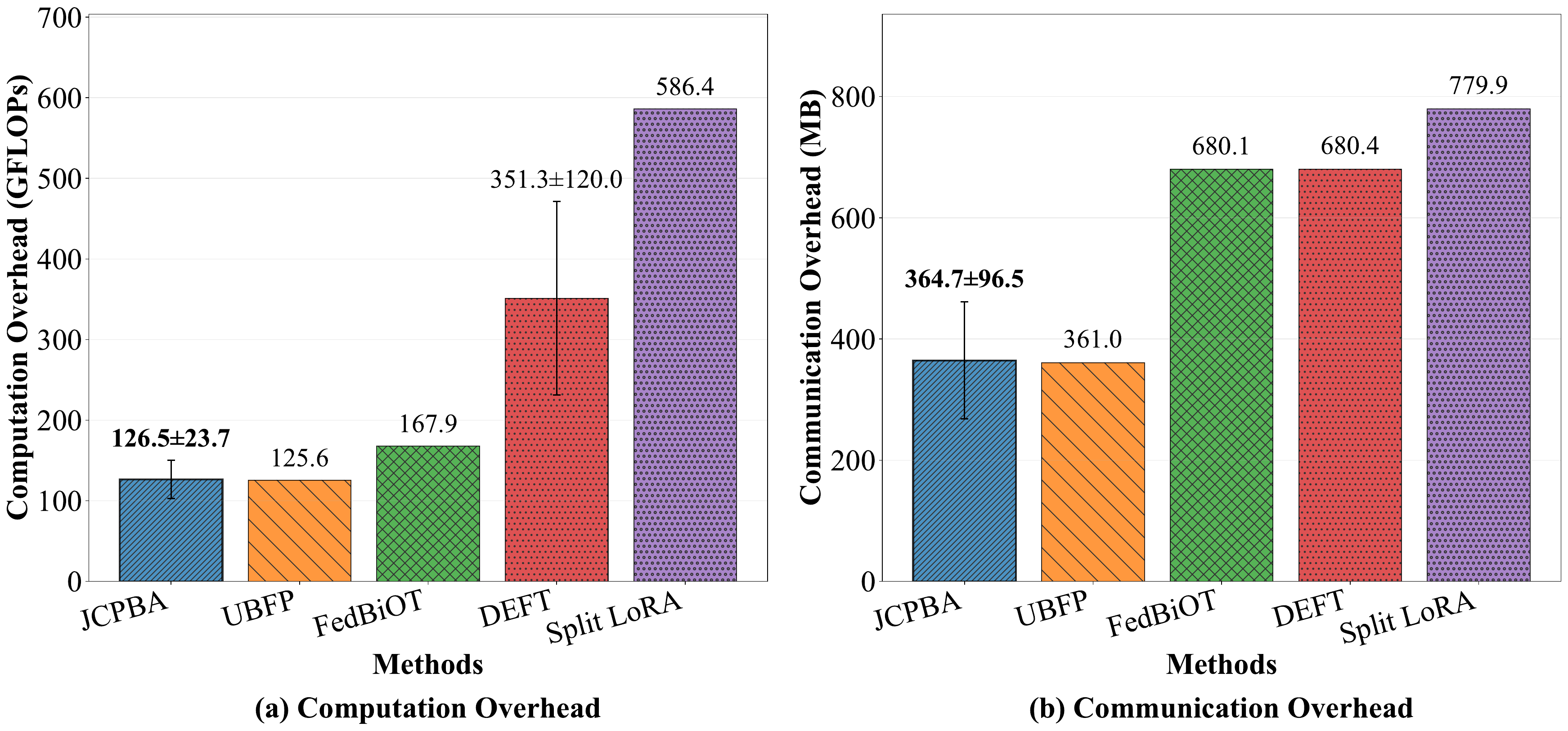}
\caption{Overhead comparison across different methods.}
\label{fig:overhead_comparison}
\end{figure}
\subsection{Results and Discussion}
\subsubsection{Fine-tuning Latency vs. Test Loss}
Figs.~\ref{fig:results}(a) and \ref{fig:results}(b) illustrate the effect of federated fine-tuning strategy on the test loss under different total fine-tuning times, with experiments conducted on the Yahoo Answers and GSM8K datasets, respectively (the x-axis denotes total fine-tuning time (s), and the y-axis denotes test loss). First, it is observed that JCPBA consistently achieves the same or even lower test loss in less time compared with all baselines on both datasets. Second, Fig.~\ref{fig:results}(a) shows that JCPBA attains a test loss of 2.2 at approximately 1,000 s in the Yahoo Answers dataset, whereas UBFP and Split-LoRA-FedFT consume more than 2,000 s to reach a comparable test loss. Meanwhile, JCPBA achieves a lower final test loss than DEFT, FedBiOT, and Split LoRA. On the GSM8K dataset, Fig.~\ref{fig:results}(b) shows that JCPBA reaches a test loss of 1.1 after 1,500 s, while UBFP needs more than 2,000 s to reach a test loss of around 1.15. In terms of the test loss, JCPBA also outperforms DEFT, FedBiOT, and Split LoRA. This observation corroborates that JCPBA efficiently reduces both computation and communication overhead while preserving and even improving model quality.
\subsubsection{Computation and Communication Overhead}
Figs.~\ref{fig:overhead_comparison}(a) and \ref{fig:overhead_comparison}(b) compare the computation and communication overhead of JCPBA with the baselines, including UBFP, FedBiOT, DEFT, and Split LoRA. First, Fig.~\ref{fig:overhead_comparison}(a) shows that JCPBA incurs a computation overhead of \(126.5 \pm 23.7\) GFLOPs (mean $\pm$ standard deviation), which is comparable with UBFP (125.6 GFLOPs) and markedly lower than FedBiOT (167.9 GFLOPs) and DEFT (\(351.3 \pm 120.0\) GFLOPs). Second, Fig.~\ref{fig:overhead_comparison}(b) shows that JCPBA incurs \(364.7 \pm 96.5\) MB of communication cost, similar to UBFP (361.0 MB) and substantially below FedBiOT (680.1 MB) and DEFT (779.9 MB). The large standard deviation of JCPBA’s overhead reflects its per-client adaptation flexibility: the dynamic pruning and bandwidth allocation effectively tailor computation and communication to specific client conditions, thereby significantly reducing the latency required to achieve the target loss.
\begin{figure}[t]
\centering
\includegraphics[width=3.5in]{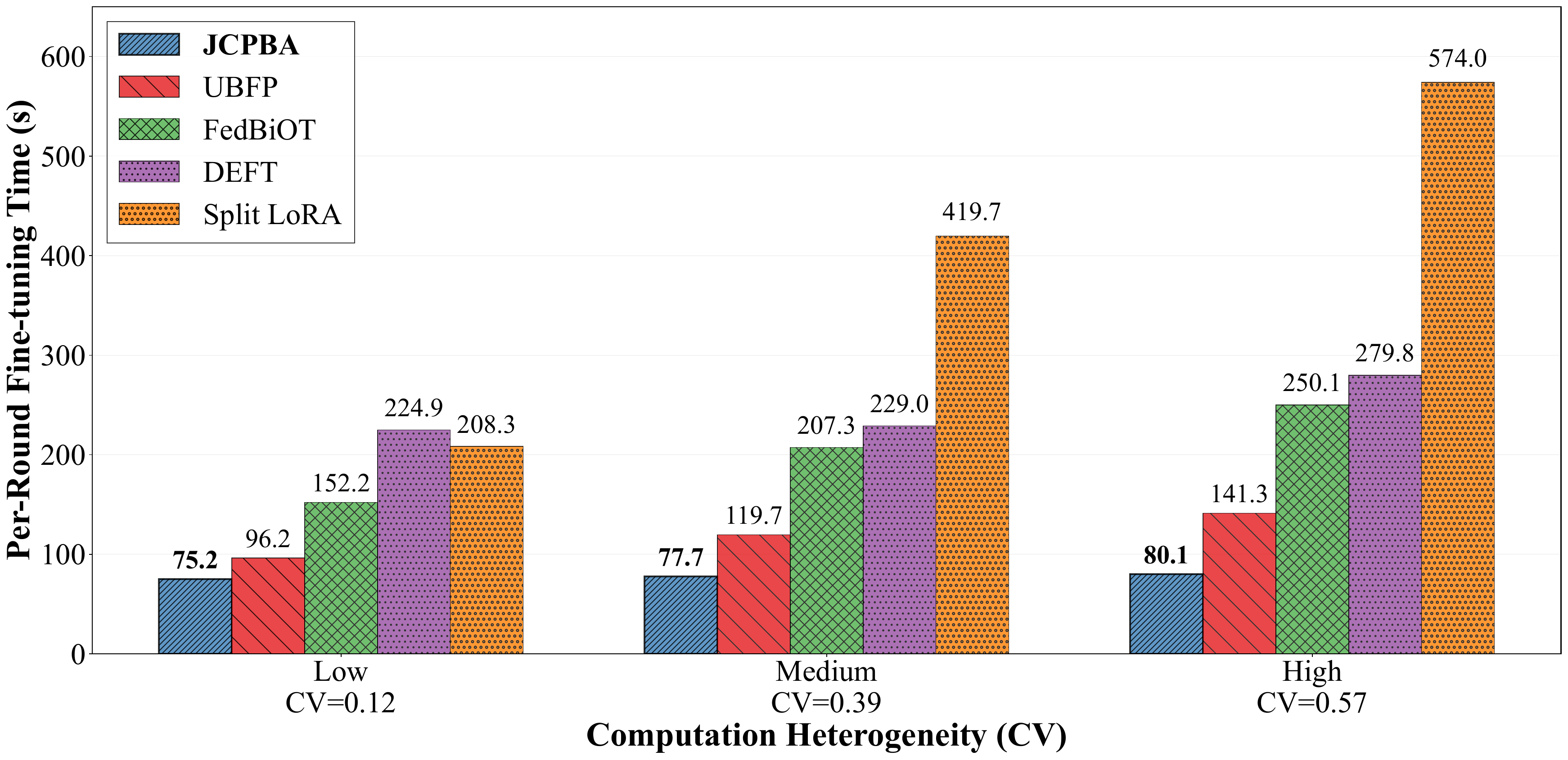}
\caption{Comparison of fine-tuning latency under different computation heterogeneity levels. }
\label{fig:latency_heterogeneity}
\end{figure}
\subsubsection{Fine-tuning Latency under Heterogeneous Computation Capacities}
Fig.~\ref{fig:latency_heterogeneity} compares the fine-tuning latency (total time to reach a fixed target test loss) of the proposed JCPBA with baselines under different levels of computation heterogeneity, where client computation factors \( \tilde{f}_k \) are sampled from three ranges (i.e., \([1, 1.5]\) at low coefficient of variation (CV) of 0.12, \([0.5, 2]\) at medium CV of 0.39, \([0.2, 2.5]\) at high CV of 0.57). First, we observe that the fine-tuning latency of all methods increases as computation heterogeneity enlarges due to straggler effects, but JCPBA exhibits the smallest latency growth across all heterogeneity levels. Specifically, JCPBA’s latency only rises from 75.2 s at CV of 0.12 to 80.1 s at CV of 0.57, i.e., an increase of about 6\%. n contrast, UBFP’s latency rises from 96.2 s to 141.3 s (i.e., around 47\% growth), FedBiOT from 152.2 s to 250.1 s (i.e., around 64\% growth), DEFT from 224.9 s to 279.8 s (i.e., around 24\% growth), and Split-LoRA-FedFT from 208.3 s to 574.0 s (i.e., around 176\% growth). These results corroborate that JCPBA effectively mitigates straggler effects and sustains efficiency under client heterogeneity.
\section{Conclusion}
We proposed the JCPBA framework for federated LLM fine-tuning over resource-constrained wireless networks. Within JCPBA, the server dynamically allocates client-specific pruning rates and bandwidth, enabling each client to perform efficient fine-tuning on the model's adapter. This approach significantly accelerates convergence while reducing computation and communication overhead. Experimental results show that JCPBA achieves faster convergence with equal or lower test loss compared with state-of-the-art baselines, reducing wall-clock time by at least 40\% and communication overhead by at least 46\%. Furthermore, JCPBA effectively mitigates straggler effects and remains robust to client computational heterogeneity, making it an efficient and scalable solution for the applications of federated LLM fine-tuning over wireless networks.

\bibliographystyle{IEEEtran}
\bibliography{ref}

\vfill

\end{document}